\PassOptionsToPackage{unicode}{hyperref}
\PassOptionsToPackage{hyphens}{url}
\documentclass[
]{article}
\usepackage{amsmath,amssymb}
\usepackage{lmodern}
\usepackage{iftex}
\ifPDFTeX
  \usepackage[T1]{fontenc}
  \usepackage[utf8]{inputenc}
  \usepackage{textcomp} % provide euro and other symbols
\else % if luatex or xetex
  \usepackage{unicode-math}
  \defaultfontfeatures{Scale=MatchLowercase}
  \defaultfontfeatures[\rmfamily]{Ligatures=TeX,Scale=1}
\fi
\IfFileExists{upquote.sty}{\usepackage{upquote}}{}
\IfFileExists{microtype.sty}{% use microtype if available
  \usepackage[]{microtype}
  \UseMicrotypeSet[protrusion]{basicmath} % disable protrusion for tt fonts
}{}
\makeatletter
\@ifundefined{KOMAClassName}{% if non-KOMA class
  \IfFileExists{parskip.sty}{%
    \usepackage{parskip}
  }{% else
    \setlength{\parindent}{0pt}
    \setlength{\parskip}{6pt plus 2pt minus 1pt}}
}{% if KOMA class
  \KOMAoptions{parskip=half}}
\makeatother
\usepackage{xcolor}
\IfFileExists{xurl.sty}{\usepackage{xurl}}{} % add URL line breaks if available
\IfFileExists{bookmark.sty}{\usepackage{bookmark}}{\usepackage{hyperref}}
\hypersetup{
  pdftitle={FACET2-S2E: Start-to-end simulations of the FACET-II beamline},
  hidelinks,
  pdfcreator={LaTeX via pandoc}}
\ifLuaTeX
  \usepackage{selnolig}  % disable illegal ligatures
\fi
\usepackage[style=numeric,sorting=none]{biblatex}
\title{FACET2-S2E: Start-to-end simulations of the FACET-II beamline}
\author{Nathan Majernik\textsuperscript{1} \and Frederick Cropp\textsuperscript{1} \and Claudio Emma\textsuperscript{1} \and Thamine Dalichaouch\textsuperscript{2}\\[0.5em]
\textsuperscript{1}SLAC National Accelerator Laboratory, USA\\
\textsuperscript{2}University of California, Los Angeles, USA}
\begin{document}
\maketitle

\hypertarget{summary}{%
\section{Summary}\label{summary}}

\href{https://github.com/slaclab/FACET2-S2E}{\texttt{FACET2-S2E}} is a Python package for start-to-end simulations of
the Facility for Advanced Accelerator Experimental Tests-II (FACET-II)
\autocite{yakimenko:2019}, a US Department of Energy National User
Facility. A kilometer-long particle accelerator creates, manipulates,
and accelerates electron beams to over 10 GeV before focusing and
compressing them to the micron-scale. These beams create extreme
electric and magnetic fields on the femtosecond timescale, uniquely
enabling research into exotic states and advanced accelerator
technology, including plasma wakefield acceleration. This software
package enables present or prospective facility users to easily run the
most common types of simulation pipelines to design experiments and
interpret results.

\hypertarget{statement-of-need}{%
\section{Statement of need}\label{statement-of-need}}

The complexity of the underlying beam dynamics in particle accelerators
generally necessitates that operators, users, and physicists who
control, interpret accelerator data and design experiments,
respectively, use simulations. Sources of complexity include the
downstream collective effects, such as coherent synchrotron radiation,
and the electron source physics. In the case of FACET-II and other
accelerators employing photoinjectors, the photocathode physics and the
space-charge dominated physics in the electron gun is also an important
consideration. FACET-II is a National User Facility focusing on the
generation of ultra-bright, extremely dense electron bunches to serve a
broad set of user experiments in applications ranging from strong-field
QED, laboratory astrophysics, plasma wakefield acceleration, and x-ray
science. Recent results include the generation of 100 kiloampere class
electron bunches using laser based beam shaping techniques
\autocite{PhysRevLett.134.085001} and the demonstration of beam
brightness transformation using a multi-stage plasma wakefield
acceleration mechanism \autocite{zhang2025}. The facility presents
specific challenges from a simulation point of view as typical
experiments generally use different simulation codes to model the beam
generation, transport through the linear accelerator including beam
shaping with lasers and collimators, and plasma simulations.

One challenge for modeling complex systems is handling multiple
simulation results in a single workflow. For workflows using multiple
simulation codes, one must ensure compatible handoffs between
simulations. For FACET-II, for reasons described in the Software Design
section, three (3) different simulation codes are used. The core
workflow uses \texttt{IMPACT-T} \autocite{qiang:2006} for beam
generation and low energy transport, \texttt{Bmad}, \texttt{Tao}, and
\texttt{PyTao} \autocite{Sagan:Bmad2006} for most of the beam transport
through the kilometer-long linear accelerator, and, optionally,
\texttt{QPAD} \autocite{li2021quasi} for particle-in-cell simulations of
the beam in plasma, and \texttt{openPMD-beamphysics}
\autocite{christopher_mayes_2025_15477845} for handling beam files.

\texttt{FACET2-S2E} is a Python package that abstracts away these
challenges from the end-user. It contains utilities, analysis notebooks,
and configuration files used to perform start-to-end (S2E) simulations
of the FACET-II particle accelerator beamline, a facility which hosts
hundreds of external users a year. Users can quickly and easily run the
most common types of simulations including parameter scans, constrained
optimization of both Twiss and multiparticle tracking objectives, and
jitter sensitivity analysis. The package seamlessly chains together
multiple codes, provides reference configurations, offers convenience
functions which mimic real-world feedback systems, and offers templates
for common development and optimization tasks, so beam physicists can
focus on the physics.

Using simulations from the control room for operators, users and
physicists is a standard use-case. FACET-II, like many accelerators,
uses EPICS \autocite{dalesio1991epics}. Another challenge of performing
control room simulations is translating between the language and units
of the underlying codes and the real-world EPICS control system. This
package provides an easy and convenient way to do simulations in the
control room. In fact, this package is readily integrated into control
room tools, both importing real values into simulations and using the
simulation optimizers to write new values to the machine. It also
abstracts away details (e.g.~specific magnet values) instead allowing
users to focus on specifying desired higher-level goals (e.g.~beam
qualities).

\hypertarget{state-of-the-field}{%
\section{State of the field}\label{state-of-the-field}}

Other software packages exist which simulate different parts of a
beamline, for example \texttt{GPT} \autocite{de1996general} and
\texttt{IMPACT} \autocite{qiang:2006} for low energy portions;
\texttt{elegant} \autocite{borland2000elegant} and \texttt{Bmad}
\autocite{Sagan:Bmad2006} for higher energy regions; or
\texttt{HiPACE++} \autocite{diederichs2022hipace} and \texttt{QPAD}
\autocite{li2021quasi} for particle-in-cell beam-plasma interactions.
These are general purpose codes which allow the design of arbitrary
beamlines but their learning curves are commensurately steep. Further,
they have different domains, requiring handoffs of the beam between
codes.

The standardization of simulations is a burgeoning sub-field within
accelerator modeling, with the OpenPMD standard
\autocite{huebl_2015_33624} defining a standard beam distribution
format, and the in-development Particle Accelerator Lattice Standard
(PALS) \autocite{pals_docs_2025} defining simulation and experimental
accelerator lattice files. Other packages aim to link together multiple
codes, like LUME \autocite{LUME}. However, this is still a general
package.

In contrast, \texttt{FACET2-S2E} is a bespoke package which models,
using several underlying codes, a single beamline which is thoughtfully
compartmentalized and abstracted, making it as frictionless as possible
for members of the FACET-II user community to perform beamline
simulations.

\hypertarget{software-design}{%
\section{Software design}\label{software-design}}

\texttt{FACET2-S2E} wraps established accelerator simulation codes into
a single workflow for the FACET-II beamline. \texttt{IMPACT-T}
\autocite{qiang:2006} generates the initial beam, \texttt{Bmad},
\texttt{Tao}, and \texttt{PyTao} \autocite{Sagan:Bmad2006} handle the
kilometer‑long transport and, optionally, \texttt{QPAD}
\autocite{li2021quasi} simulates plasma wakefield sections. \texttt{IMPACT-T} is
a good choice for modeling photoinjectors and the electron source
physics. Bmad was chosen as a fast, but accurate solver outside that
regime. QPAD was chosen as a fast, highly efficient plasma simulation
tool due to its hybrid algorithm which combines both the quasi-3D
Fourier mode decomposition \autocite{lifschitz09} and the quasi-static
approximation \autocite{mora97}. The quasi-3D algorithm decomposes
fields and sources into azimuthal fourier modes on a cylindrical grid,
thereby reducing the algorithmic complexity to that of a 2D code while
still resolving 3D asymmetries. Under the quasi-static approximation,
the particle beam evolves slowly over the plasma response period,
thereby permitting large timesteps and orders of magnitude computational
speedup over conventional PIC codes. Beam files are exchanged via the
openPMD format. Users interact with high‑level functions that abstract
away the multiple underlying codes, as well as translating the units and
language to match that of the EPICS variable names and units present in
the control room. Configuration files centralise facility‑specific
defaults and can be overridden without touching the core code, while
helpers support parameter scans, optimization, and jitter studies. The
result is a simplified toolkit tailored to a single beamline.

\hypertarget{research-impact-statement}{%
\section{Research impact statement}\label{research-impact-statement}}

While under development, this package has been used to design
experiments, support real-time execution of experiments from the control
room, and aid in the interpretation of experimental results. Such
experiments include those published as \autocite{PhysRevLett.134.085001}
and \autocite{zhang2025}.

\hypertarget{ai-usage-disclosure}{%
\section{AI usage disclosure}\label{ai-usage-disclosure}}

Generative AI, specifically LLMs from OpenAI and Anthropic, were used in
software and documentation creation. All AI generated content has been
reviewed by humans before being included in the package.

\hypertarget{acknowledgements}{%
\section{Acknowledgements}\label{acknowledgements}}

This work was supported by the U.S. Department of Energy under DOE
Contract No.~DE-AC02-76SF00515 and the Consortium for Advanced Modeling
of Particle Accelerators (CAMPA) funded under DOE Contract
No.~DE-AC02-05CH11231. The authors thank M. Ehrlichman and C. Mayes for
sharing their Bmad expertise and D. Cesar for his floorplan plotting
function.

\printbibliography[title=References]

\end{document}